\documentclass[twoside,fleqn]{article}
\usepackage{espcrc2}
\usepackage{theorem}
\usepackage{epsfig}
\usepackage{graphicx}

\theoremstyle{break}    
\theoremstyle{plain}    
\theoremstyle{plain}    
\theoremstyle{plain}    
{\theorembodyfont{\rmfamily}     }
{\theorembodyfont{\rmfamily}     }

\def\gsim{{\mathrel{\raise2pt\hbox to 8pt{\raise -5pt\hbox{$\sim$}\hss{$>$}}}}}
\def\rsim{{\mathrel{\raise2pt\hbox to 8pt{\raise -5pt\hbox{$\sim$}\hss{$>$}}}}}
\def\lsim{{\mathrel{\raise2pt\hbox to 8pt{\raise -5pt\hbox{$\sim$}\hss{$<$}}}}}

\begin{document}

\title{
       \begin{flushright}\normalsize
	    \vskip -0.9 cm
            SNUTP-03-021
       \end{flushright}
	\vskip -0.4 cm
Penguin diagrams for the HYP staggered fermions\thanks{ Presented by
K.~Choi.  Research supported in part by BK21, by the SNU foundation \&
Overhead Research fund, by KRF contract KRF-2002-003-C00033 and by
KOSEF contract R01-2003-000-10229-0.}  }
\author{
        Keunsu~Choi\address[SNU]{School of Physics, 
		Seoul National University,
		Seoul, 151-747, South Korea}
	and 
	Weonjong~Lee\addressmark[SNU]
}
\begin{abstract}
We present results of the one-loop corrections originating from
the penguin diagrams for the improved staggered fermion operators
constructed using various fat links such as Fat7, Fat7+Lepage,
$\overline{\rm Fat7}$, HYP (I) and HYP (II). The main results
include the diagonal/off-diagonal mixing coefficients and
the matching formula between the continuum and lattice operators.
\end{abstract}

\maketitle

\section{INTRODUCTION}\label{sec:intr}
The low energy effective Hamiltonian of the standard model includes
$\Delta S = 1$ four-fermion operators with corresponding Wilson
coefficients, which contains all the short-distance physics.
The low energy effects of the electroweak and strong interactions can
be expressed in terms of matrix elements of the four-fermion operators
between hadronic states.
Lattice QCD is well-suited to calculate these matrix elements
non-perturbatively at low energy.
One essential step in using lattice QCD is to find the relationship
between the continuum and lattice operators, which is often called
``matching formula''.
There are two classes of Feynman diagrams at the one-loop level:
(1) current-current diagrams and (2) penguin diagrams.
At the one loop level, it is possible to treat the penguin
contribution and the current-current contribution separately.
In the case of the current-current diagrams, the matching formula
at the one-loop level is given in \cite{ref:wlee:1}.
Here, we focus on penguin diagrams in which one of the quarks in the
four-fermion operator is contracted with one of the anti-quarks to
form a closed loop.
Hence, the main goal is to calculate the penguin diagrams for improved
staggered operators constructed using various fat links and to
provide the corresponding matching formula.
Here, we adopt the same notation and Feynman rules outlined in
\cite{ref:wlee:1}.
\section{PENGUIN DIAGRAMS}
\begin{figure}[t]
\epsfig{file=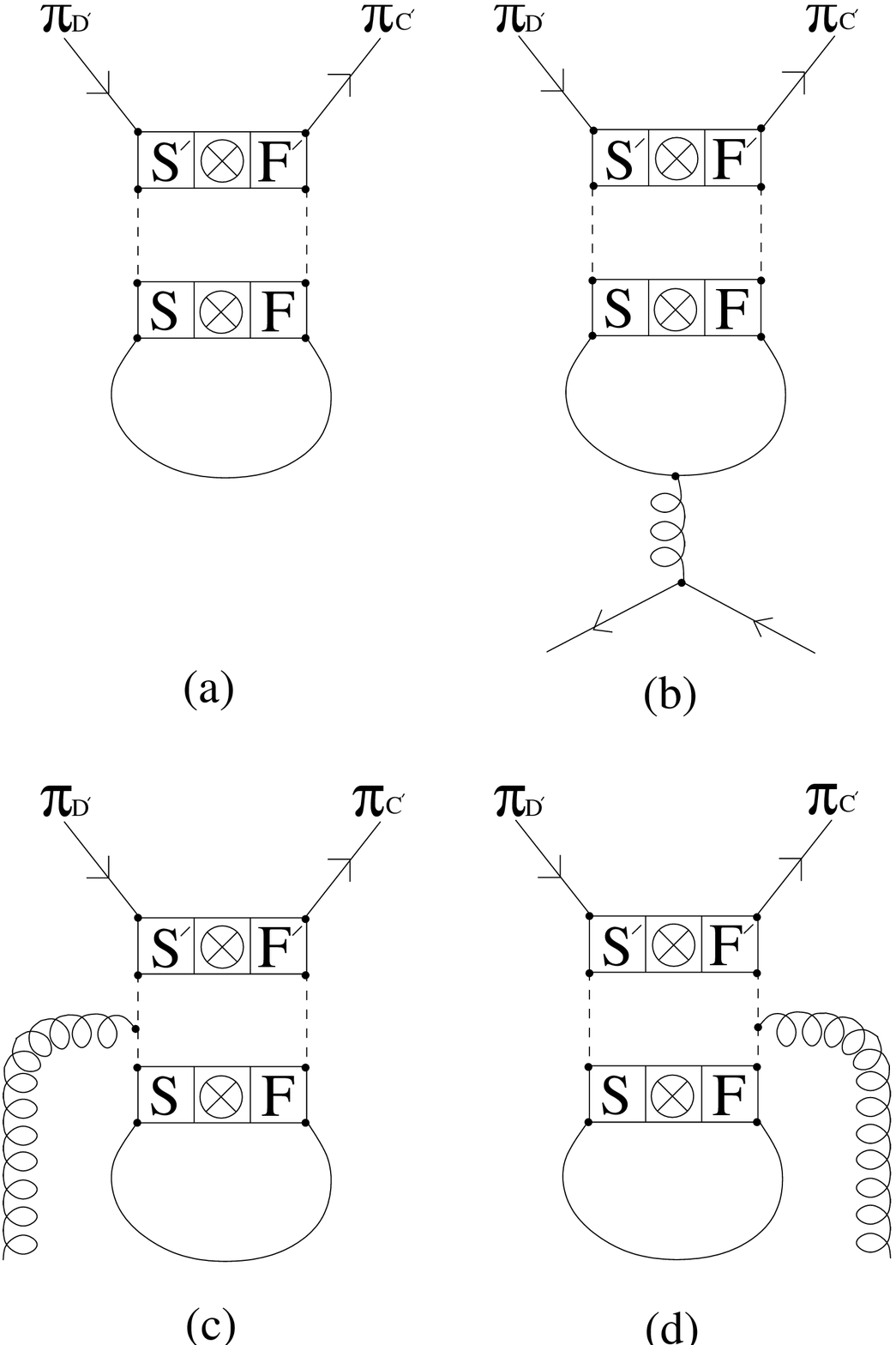, width=16pc}
\vspace*{-5mm}
\caption{Penguin diagrams.}
\label{fig:1}
\end{figure}

\begin{figure}[t]
\epsfig{file=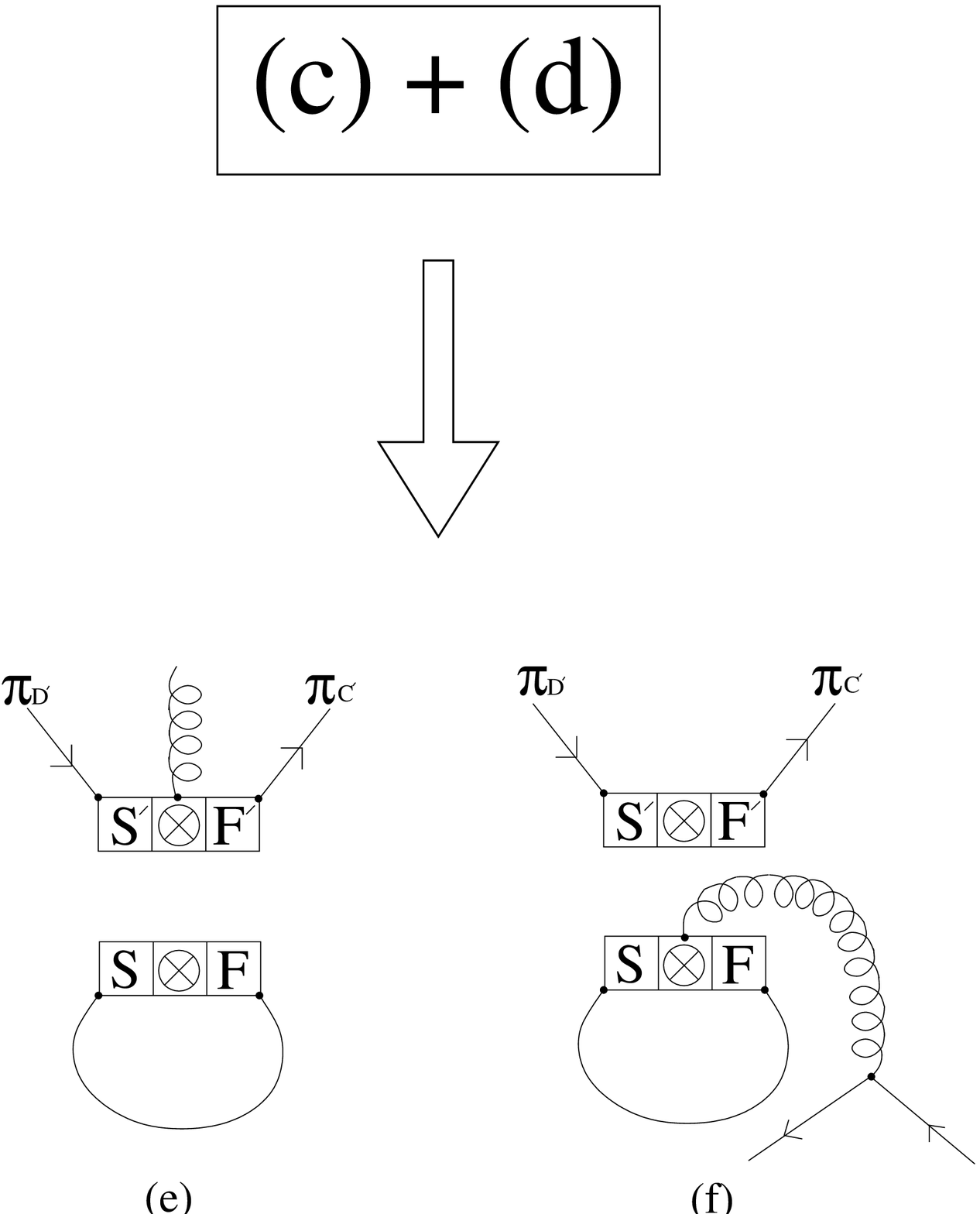, width=16pc}
\vspace*{-5mm}
\caption{Diagram identity.}
\label{fig:2}
\end{figure}

\begin{figure}[t]
\begin{center}
\epsfig{file=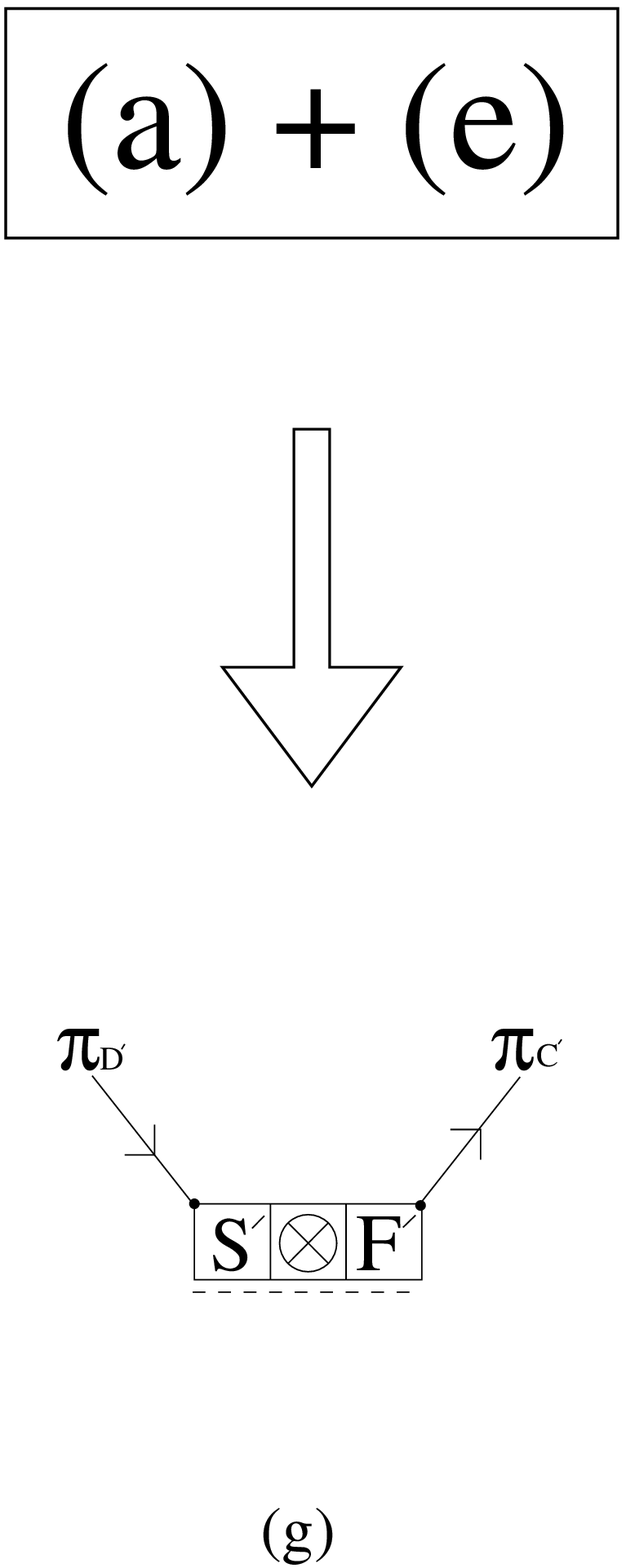, width=8pc}
\end{center}
\vspace*{-5mm}
\caption{Bilinear operator mixing.}
\label{fig:3}
\end{figure}

\begin{figure}[t]
\epsfig{file=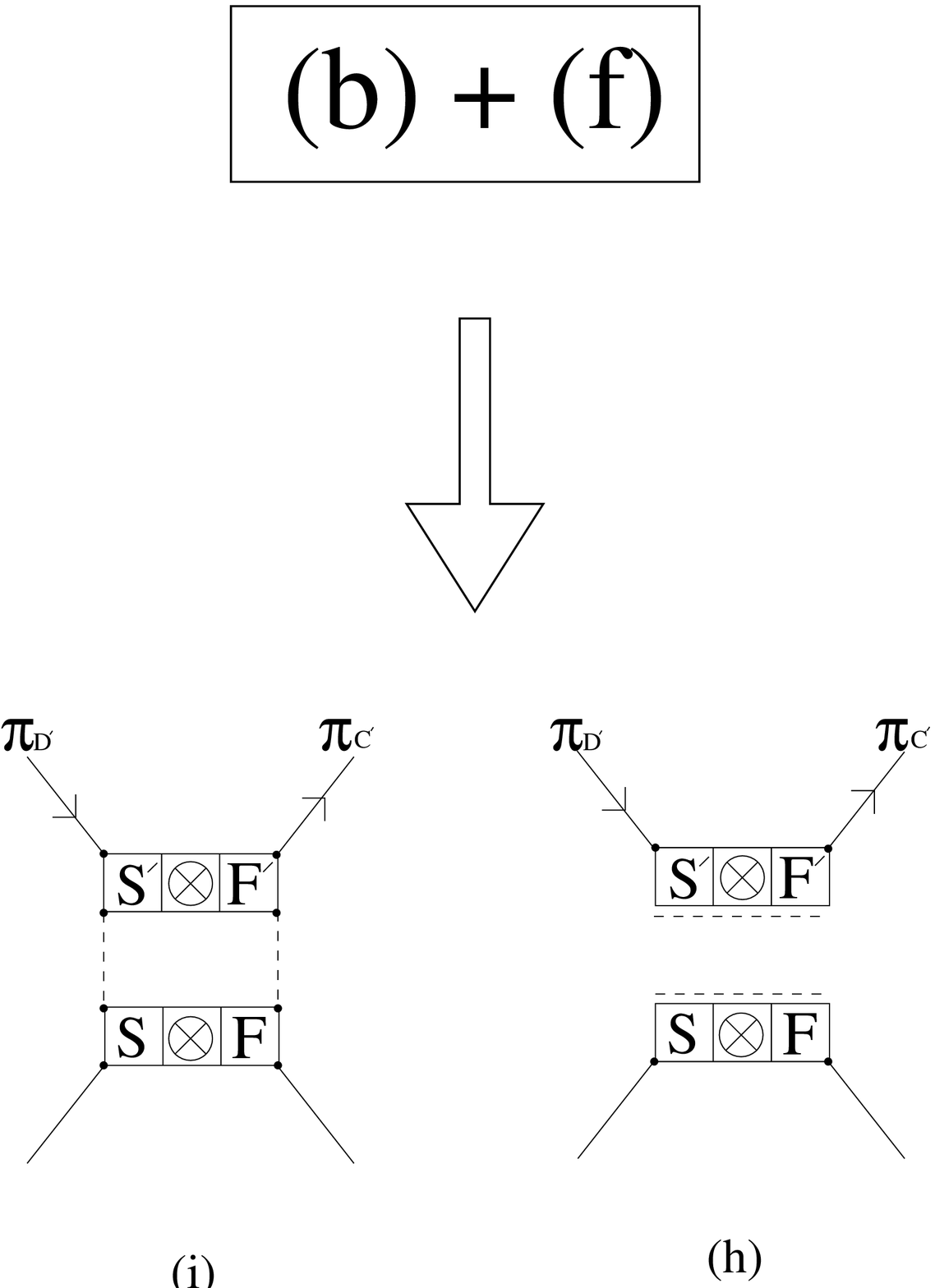, width=16pc}
\vspace*{-5mm}
\caption{Four-fermion operator mixing.}
\label{fig:4}
\end{figure}

Here, we study penguin diagrams. 
On the lattice, the gauge non-invariant four fermion operators such as
Landau gauge operators mix with lower dimension operators, which are
gauge non-invariant \cite{ref:sharpe:1}.
It is required to subtract these contributions non-perturbatively.
However, it is significantly harder to extract the divergent mixing
coefficients in a completely non-perturbative way.
Therefore, it is impractical to use gauge non-invariant operators
for the numerical study of the CP violations. 
Hence, it is prerequisite to use gauge invariant operators in order
to avoid unwanted mixing with lower dimension operators.
For this reason, we choose gauge invariant operators in this study. 
In the staggered fermion formalism, there are four penguin diagrams 
at the one loop level as shown in Fig.~\ref{fig:1}.
These diagrams allow mixing with lower dimension operators as well as
four fermion operators of the same dimension or higher.
The mixing coefficients with lower dimension operators are divergent
({\em i.e.} proportional to inverse power of the lattice spacing).
The perturbation is, however, not reliable with divergent coefficients.
Hence, we must use non-perturbative method to determine them and 
subtract away the lower dimension operators.
In the case of mixing with operators of the same dimension, the
perturbation is expected to be reliable as long as the size of
one-loop correction is small enough, which can be achieved
by using improved staggered fermions.
In Fig.~\ref{fig:1}, diagrams (a) and (b) have their correspondence in
the continuum and diagrams (c) and (d) are pure lattice artifacts.
However, diagrams (c) and (d) play an essential role to keep the gauge
invariance.
Basically, the contribution from diagrams (c) and (d) can be
re-expressed as a sum of diagrams (e) and (f) as shown in
Fig.~\ref{fig:2}.
The first key point is that the sum of diagrams (a) and (e) 
generates bilinear operators in a gauge invariant form.
The main key point is that the contributions from diagrams (b) and (f)
,as shown in Fig.~\ref{fig:4}, leads to four fermion operators of our
interests in a gauge invariant form.
The details of bilinear mixing (diagrams (a) and (e)) in
Fig.~\ref{fig:3} will be presented in \cite{ref:wlee:2} and here we
skip them.
Here we focus on the diagrams (b) and (f) and present the final
result.
The final result is 
\begin{eqnarray}
G_{(b+f)} &=& \bigg( - \frac{1}{N_f} \bigg) 
  \frac{g^2}{ (4\pi)^2 } \bigg(\sum_I T^I_{ab} T^I_{cd} \bigg) I_c 
\nonumber \\
 & & \cdot \sum_{\mu} 
\overline{\overline{ (\gamma_{S'} \otimes \xi_{F'}) }}_{C'D'} 
\overline{\overline{ (\gamma_{\mu} \otimes 1 ) }}_{CD} 
\nonumber \\
& & \cdot  \delta_{S,\mu} \delta_{F,1} \Big[ h_{\mu\mu}(k) \Big]^2 
\label{eq:b+f:1}
\end{eqnarray}
where $k = q-p$ is strictly on shell.
\begin{eqnarray}
I_c &=& \frac{16}{3} \bigg( 
-\ln( 4 m^2 a^2 ) - \gamma_E + F_{0000} \bigg) - 9.5147 
\nonumber \\
& & + {\cal O}(m^2 a^2)
\end{eqnarray}
$I_c$ is also given in \cite{ref:sharpe:1}.
The details of deriving Eq.~(\ref{eq:b+f:1}) will be
presented in \cite{ref:wlee:2}.
From Eq.~(\ref{eq:b+f:1}) we can derive the following
theorem:
\noindent {\bf Theorem 1 (Equivalence)} \\
{\em At the one loop level, the diagonal mixing coefficients of
penguin diagrams are identical between (a) the unimproved (naive)
staggered operators constructed using the thin links and (b) the
improved staggered operators constructed using the fat links such as
HYP (I), HYP (II), Fat7, Fat7+Lepage, and $\overline{\rm
Fat7}$}.\footnote{Note that AsqTad is NOT included on the list. In
this case, by construction the operators are made of the fat links
which are not the same as those used in the action due to the Naik
term. In addition, the choice of the fat links are open and not
unique.}
The details on the proof of this theorem will be given in
\cite{ref:wlee:2}.
By construction, gluons carrying a momentum close to $k \sim \pi/a$
are physical in staggered fermions and lead to taste changing
interactions, which is a pure lattice artifact.
In the case of unimproved staggered fermions, it is allowed to mix
with wrong taste ($\ne 1$) and the mixing coefficient is substantial.
In contrast, in the case of improved staggered fermions using fat
links of our interest such as Fat7, $\overline{\rm Fat7}$ and HYP
(II), the off-diagonal mixing with wrong taste vanishes and is absent.
In the case of the improvement using HYP (I) and Fat7 + Lepage, the
off-diagonal mixing with wrong taste is significantly suppressed.
The details of this off-diagonal mixing will be given in
\cite{ref:wlee:2}.
In summary, the diagonal mixing occurs only when the original operator
has the spin and taste structure of $S=\mu$ and $F=1$ regardless of
that of the spectator bilinear.
The diagonal mixing coefficient is identical between the unimproved
staggered operators and the improved staggered operators constructed
using fat links such as Fat7, Fat7+Lepage, $\overline{\rm Fat7}$, HYP
(I) and HYP (II).
This is a direct consequence of the fact that the contribution from
the improvement changes only the mixing with higher dimension
operators and off-diagonal mixing, which are unphysical.

The result of this paper, combined with that of \cite{ref:wlee:1}
provides a complete set of one-loop matching formula.
%
%
%
%
%

%
%

\end{document}